# Gate-tunable spin Hall effect in trilayer graphene/group-IV monochalcogenide van der Waals heterostructures

*Haozhe Yang[1,2*], Zhendong Chi[1], Garen Avedissian[1], Eoin Dolan[1], Muthumalai Karuppasamy[3], Beatriz Martín-García[1,4], Marco Gobbi[4,5], Zdenek Sofer[3], Luis E. Hueso[1,4], Fèlix Casanova[1,4*]*

Haozhe Yang, Zhendong Chi, Garen Avedissian, Eoin Dolan, Beatriz Martín-García, Luis E. Hueso, Fèlix Casanova
CIC nanoGUNE BRTA,
Donostia-San Sebastian 20018, Basque Country, Spain.
E-mail: f.casanova@nanogune.eu

Haozhe Yang
Fert Beijing Institute, MIIT Key Laboratory of Spintronics, School of Integrated Circuit Science and Engineering,
Beihang University,
Beijing 100191, China.
E-mail: haozheyang.nanogune@gmail.com

Muthumalai Karuppasamy, Zdenek Sofer
Department of Inorganic Chemistry,
University of Chemistry and Technology Prague,
Prague 16000, Czech Republic

Beatriz Martín-García, Marco Gobbi, Luis E. Hueso, Fèlix Casanova
IKERBASQUE, Basque Foundation for Science,
Bilbao 48009, Basque Country, Spain.

Marco Gobbi
Centro de Física de Materiales (CSIC-EHU/UPV) and Materials Physics Center (MPC),
Donostia-San Sebastian 20018, Basque Country, Spain

Keywords: Proximity effect, graphene, group-IV monochalcogenide, spin-charge interconversion

Spintronic devices require materials that facilitate effective spin transport, generation, and detection. In this regard, graphene emerges as an ideal candidate for long-distance spin transport owing to its minimal spin-orbit coupling, which, however, limits its capacity for effective spin manipulation. This problem can be overcome by putting spin-orbit coupling materials in close contact to graphene leading to spin-orbit proximity and, consequently, efficient spin-to-charge conversion through mechanisms such as the spin Hall effect. Here, we report and quantify the gate-dependent spin Hall effect in trilayer graphene proximitized with tin sulfide (SnS), a group-IV monochalcogenide which has recently been predicted to be a viable alternative to transition-metal dichalcogenides for inducing strong spin-orbit coupling in graphene. The spin Hall angle exhibits a maximum around the charge neutrality point of graphene up to room temperature. Our findings expand the library of materials that induce spin-orbit coupling in graphene to a new class, group-IV monochalcogenides, thereby highlighting the potential of two-dimensional materials to pave the way for the development of innovative spin-based devices and future technological applications.

## 1. Introduction

Van der Waals (vdWs) heterostructures provide an ideal platform to induce and control various physical phenomena[1,2]. In the field of spintronics, vdW materials have been proven to exhibit unique properties for spin generation, transport and detection[3]. Graphene, a light material with weak spin-orbit coupling (SOC), has been considered as an ideal spin transport channel due to its long spin diffusion length[4,5]. On the other hand, the weak SOC of pristine graphene leads to difficulties in generating and detecting spins, thus limiting its potential as a platform for spin manipulation. One possible way to overcome this bottleneck is functionalizing graphene by proximity with materials with strong SOC[6–19]. The proximity effect can transform a given material from its pristine state by giving it some of the properties of the neighboring material. VdW heterostructures provide a suitable playground to design proximitized materials because of the short-range interfacial interaction. Graphene-based vdW heterostructures have attracted much attention as graphene shows a sizeable spin Hall angle ($\theta_{SH}$) gained from the neighboring material while keeping a long spin diffusion length ($\lambda_s$) compared to the conventional SOC materials such as heavy metals, leading to a large spin-charge interconversion (SCI) efficiency ($\theta_{SH}\lambda_s$)[20]. This is important for spintronic applications such as MESO logic[21,22] or potential applications for magnetic random access memory[23,24].

One representative example of SOC enhancement occurs when graphene is in proximity with a transition metal dichalcogenide (TMD). The graphene/TMD heterostructure was predicted to show the spin Hall effect (SHE)[25–28] and Rashba-Edelstein effect[27–29], and this was later confirmed experimentally[14,16,18,19]. Another example are the metal oxides, including heavy metal oxides[15] and light metal oxides[30], which have also been experimentally demonstrated to induce the SHE in graphene. Additionally, it has been observed that the magnetic exchange coupling can also be imprinted into graphene through proximity effect [31–34], which could induce anomalous Hall effect[31–33,35]. Recently, another compound family, the group-IV monochalcogenides MX (M=Sn, Ge; X=S, Se, Te) have attracted much attention due to their unique physical properties such as selective valley excitations[36,37], valley Hall effects[38], or ferroelectric behavior[39–41]. Orthorhombic compounds like tin sulfide (SnS) display a layered structure, making it feasible to create vdW heterostructures with other layered materials. However, directly employing SnS in spintronic devices presents challenges due to its high resistance, which makes electrical spin injection from a metallic ferromagnet difficult, because of the phenomenon known as conductivity mismatch[42,43]. A solution to this problem involves forming a proximitized heterostructure with graphene. Indeed, a recent prediction suggests that group-IV monochalcogenides can transfer SOC to graphene through the proximity effect[44]. This opens up the potential for using graphene/group-IV monochalcogenide heterostructures in spintronic applications.

In this paper, we report the first experimental observation of the SHE in trilayer graphene/SnS heterostructure. This vdW heterostructure is nanofabricated into a lateral spin valve and the SCI is measured with non-local spin precession experiments, which gives the unambiguous determination of the SHE[14,16,18,30,45,46]. We find that the graphene gains SOC, exhibiting SHE up to room temperature. The output of the SCI, as well as the spin Hall angle ($\theta_{SH}$), is gate tunable and exhibits a maximum when the Fermi level is around the charge neutrality point (CNP) of the graphene. Our work shows the potential of group-IV monochalcogenides in spintronic applications that exploit 2D materials.

## 2. Results and discussion

The schematic of the graphene/SnS heterostructure is shown in Figure 1(a). The SnS flake is mechanically exfoliated in an Ar glovebox, and immediately dry stamped on pre-exfoliated trilayer graphene, which has been observed to have a longer $\lambda_s$ compared to monolayer graphene on SiO$_2$/Si substrate[47]. Next, we fabricate a device that can detect the SCI by combining a graphene Hall bar (with SnS on top of the cross-junction and arms of pristine graphene) with ferromagnetic (FM) electrodes (see Figure 1(b)). The non-magnetic contacts are attached to both the graphene arms (N1 to N4) and pristine SnS (N5 and N6). Adjacent FM electrodes (F1 and F2) are used to calibrate the spin injection efficiency of the FM contacts and the spin transport properties of the pristine graphene, required to quantify the SCI efficiency. Details of the device fabrication are given in the Methods section.

The structural properties of SnS and graphene after device fabrication are characterized by Raman spectroscopy. In Figure 1(c), we observe a clear separation between the A$_g$ and B$_{3g}$ peaks of SnS, indicating the well-preserved structure of the exfoliated SnS flake. The 2D peak of the graphene is shown in Figure 1(d), with the fitting indicating a trilayer thickness[48]. The low intensity of the D peak in the inset indicates no evident damage to the graphene during fabrication.

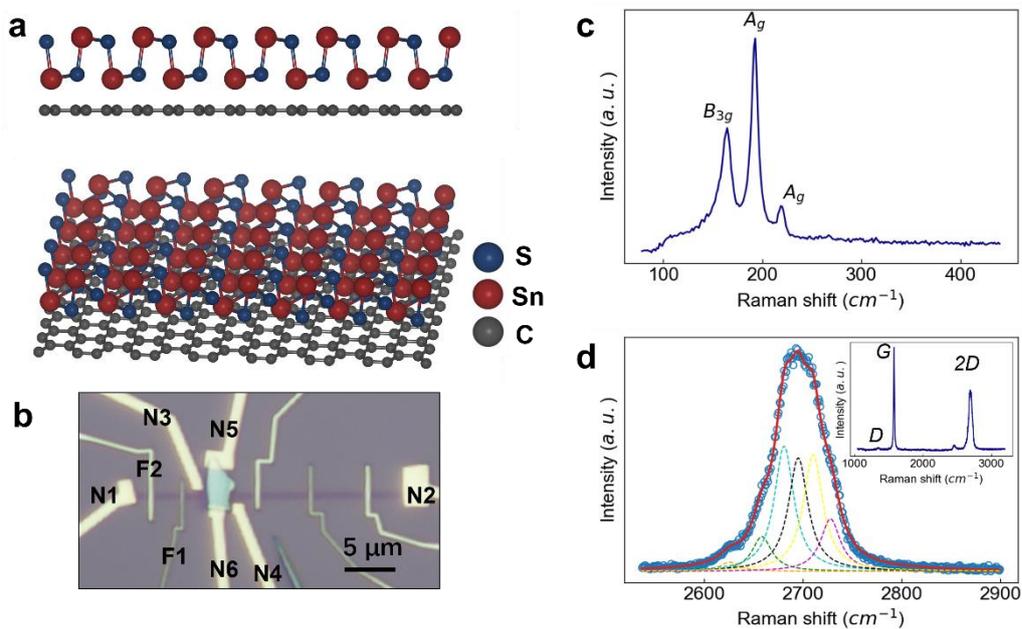

**Figure 1**. Schematic of graphene/SnS heterostructure and characterization. (a) Crystal structure sketch of the SnS/graphene heterostructure. The blue, red and gray spheres represent sulfur, tin and carbon atoms, respectively. Only one layer of the SnS and graphene is plotted for clarity. (b) Optical image of the device, where the nonmagnetic contact (N1 to N4) to graphene,

nonmagnetic contact (N5 and N6) to SnS, and magnetic electrodes (F1 and F2) to graphene are indicated. (c) Representative Raman spectrum for the SnS flake after fabrication and electrical measurement of the device, with the three Raman modes labeled. (d) 2D peak of the Raman spectrum of the pristine graphene after fabrication and electrical measurement of the device. A fit using 6 Lorentzian functions with Full Width Half Maximum of 24 cm$^{-1}$ is performed, evidencing it is trilayer graphene. Inset: Raman spectrum of the pristine graphene, with the 3 main peaks labeled.

After confirming the structural properties of the graphene and SnS, we study the charge transport properties with the measurement configuration shown in Figure 2(a). Here, only one layer of the SnS and graphene is plotted in the schematic for clarity. For SnS, a constant source-drain voltage of 0.1 V is applied between N5 and N6, with the source-drain current ($I_{sd}$) measured as a function of temperature as shown in Figure 2(b). A clear semiconducting behavior can be observed, in which the conductivity decreases from 1.4 µS at 300 K to 60 pS at 100 K. The conductivity as a function of the back gate voltage ($V_g$) measured at 300 K is shown in Figure 2(c), indicating a p-type semiconducting behavior of SnS, which is consistent with previous reports[40,49]. The resistance of the proximitized graphene, measured with a two-point configuration between N3 and N4 as a function of $V_g$, is shown in Figure 2(d). Here, the resistance exhibits a representative Dirac material feature for all temperatures, with the resistance maximum corresponding to the charge neutrality point (CNP). The CNP at 300 K shows a slight shift towards negative $V_g$, indicating possible charge transfer between SnS and graphene. In this study, the resistance of SnS is more than three orders of magnitude larger than graphene at room temperature, and even larger at lower temperature because of the semiconducting behavior of SnS. This indicates that the charge transport, along with the spin transport, mainly happens in graphene at all temperatures.

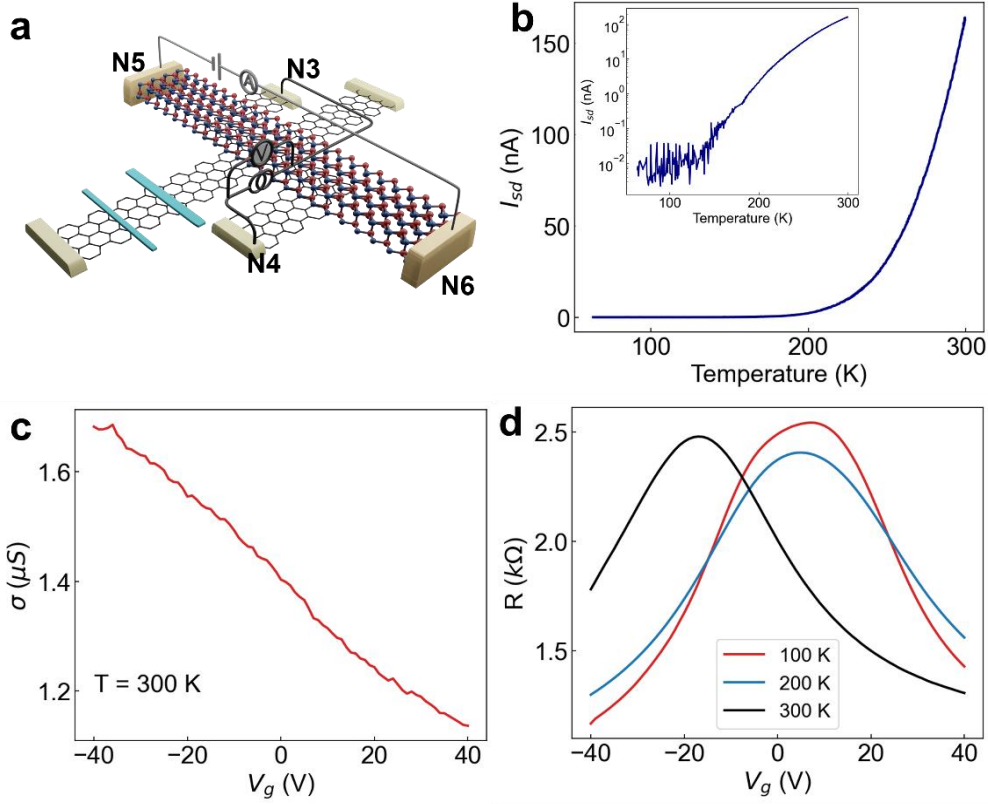

**Figure 2**. Charge transport properties of graphene/SnS heterostructure (a) Schematic of the device with the electrical configurations for the charge transport measurements. (b) Source-drain current of the SnS flake as a function of temperature by applying constant source-drain voltage (0.1 V) through N5 and N6 at $V_g = 0$. The inset shows the same data with semi-log scale. (c) Conductance of the SnS flake, measured between N5 and N6, as a function of $V_g$ at 300 K. (d) Two-point resistance of the proximitized graphene measured between N3 and N4 at different temperatures.

After obtaining the charge transport properties of the SnS and graphene, we characterize the spin transport of the graphene, with the measurement configuration shown in Figure 3(a). Here, we use two FM electrodes as the spin injector (F1) and detector (F2) to study the spin transport properties of the pristine graphene, by applying a charge current $I_c$ from F1 to N2 and detecting the nonlocal voltage $V_{NL}$ between F2 and N1. We then normalize $V_{NL}$ by $I_c$ to obtain a nonlocal resistance $R_{NL}$. First, the relative alignment of the magnetization of F1 and F2 electrodes (parallel or antiparallel) is set with an external magnetic field ($B_y$) because the different widths of F1 and F2 yield different coercivities. Subsequently, an in-plane magnetic field ($B_x$) along the hard-axis of FM contacts is swept from zero until full saturation. At low fields, spins in graphene precess in the $y-z$ plane, and $R_{NL}$ exhibits a symmetric Hanle precession behavior. One representative result measured at 100 K with $V_g = 0$ V is shown in Figure 3(b), with the data measured at 200 K and 300 K shown in Supporting Information Note 2. The pure spin precession signal $\Delta R_{NL}$ plotted in Figure 3(c) is obtained by subtracting the $R_{NL}$ curves between the parallel ($R_{NL}^P$) and antiparallel ($R_{NL}^{AP}$) configuration. $\Delta R_{NL}$ is fitted with the numerical solution of a Bloch function to extract the spin polarization of the FM and the spin diffusion constant ($D_s$) and spin lifetime ($\tau_s$) of the pristine graphene. The obtained spin diffusion length (calculated as $\lambda_s = \sqrt{D_s \tau_s}$) and $\tau_s$ at different $V_g$ and temperatures (100, 200, and 300 K) are shown in Figure 3(d). $\tau_s$ increases with decreasing temperature, whereas $\lambda_s$ remains almost constant at all measured temperatures and backgate voltages for the trilayer graphene. In the following section, we use these values to estimate $\theta_{SH}$ and $\tau_s$ for the proximitized graphene.

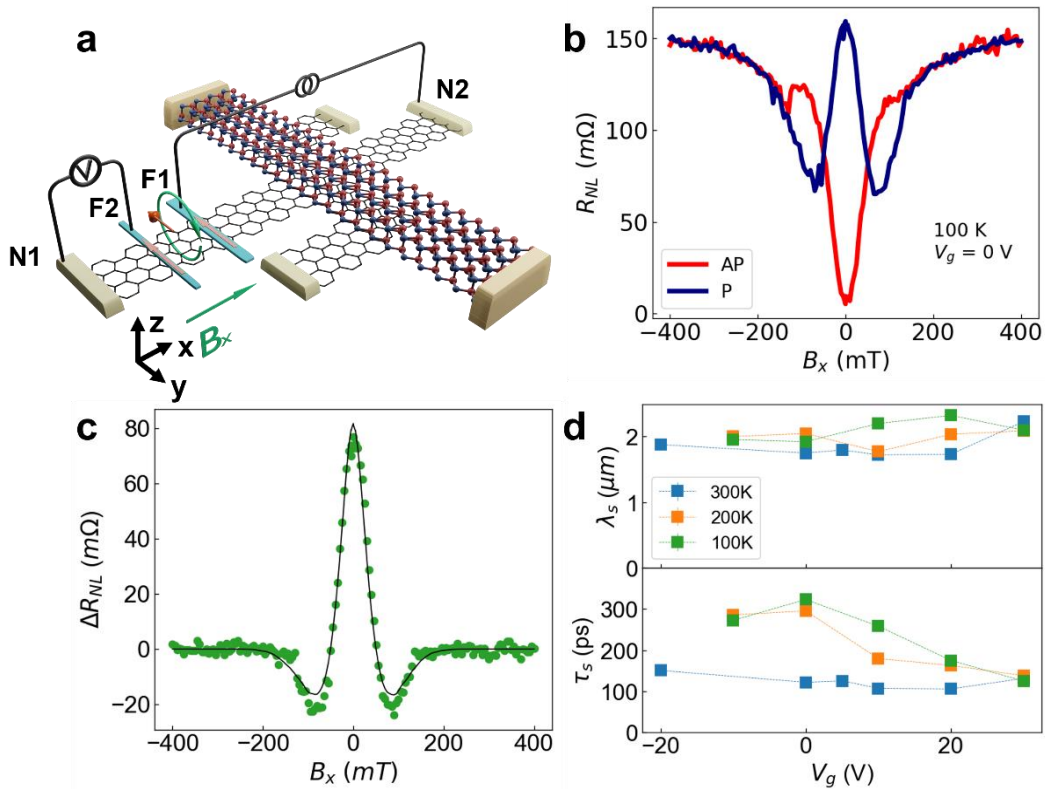

**Figure 3**. Spin transport properties of the pristine graphene. (a) Sketch of non-local Hanle precession measurement, with an in-plane magnetic field applied to induce precession of the *y*-polarized spins injected from the FM electrode into the graphene. (b) Hanle precession measurement measured at 100 K and $V_g = 0$ V, with F1 and F2 electrodes set in a parallel (P, blue line) and antiparallel (AP, red line) configuration. (c) Net symmetric Hanle precession signal extracted from the two curves in (b) and calculating $\Delta R_{NL} = (R_{NL}^P - R_{NL}^{AP})/2$. The gray

solid line is a fit of the data to the solution of the Bloch equation. (d) Spin diffusion length (upper panel) and spin lifetime (lower panel) as a function of backgate voltage measured at 100 K (green), 200 K (orange), and 300 K (blue) The error bar, calculated from the fitting error of the symmetric Hanle precession, is smaller than the size of the symbols.

Next, we perform the spin-charge interconversion experiments, with the measurement configuration shown in Figure 4(a). A charge current ($I_c$) is applied from F1 to N1. Due to electrical spin injection, a spin current is generated at the interface between F1 and graphene, with its polarization parallel to the FM magnetization along *y*-direction. By applying an in-plane magnetic field ($B_x$), the injected *y*-spins precess in the *y*-*z* plane, causing the diffusing spins to develop a *z*-component at finite magnetic fields. When the spin current reaches the proximitized region, the *z*-spins are converted into a transverse charge current through the inverse SHE, resulting in a non-local voltage ($V_{NL}$) between N3 and N4. Again, we normalize $V_{NL}$ by $I_c$ to obtain a non-local resistance ($R_{NL}$). Note that spin-to-charge conversion and charge-to-spin conversion can be measured by swapping the voltage and current terminals, following Onsager reciprocity[12,30]. We will refer to both cases as SCI for simplicity. This process results in an antisymmetric spin precession curve with a maximum and a minimum at a certain $\pm B_x$ value. When reversing the magnetization of F1, the antisymmetric Hanle curve is reversed because the detector senses the opposite *y*-spin component. To obtain the two curves, we initialize the F1 magnetization with a $B_y$ field, then sweep $B_x$ from zero until saturation of the electrode. This operation is repeated for the two F1 magnetizations and $B_x$ polarities. The non-local curves measured at $V_g = 0$ V and different temperatures (100, 200, and 300 K) are plotted in Figure 4(b), showing clear antisymmetric spin precession behavior. Here, the baseline is shifted to zero to compare the curves at different temperatures. We can observe in Figure 4(b) that the amplitude, noted as $\Delta R_{SCI}$, decreases with increasing the temperature. This has also been observed in graphene/WSe$_2$ (Ref.[14]) and graphene/BiO$_x$ heterostructures[15], which could be attributed to the weakening of the proximity effect when increasing the temperature. Nevertheless, a distinct SCI can still be observed at room temperature, suggesting the viability of using graphene/SnS heterostructures for SCI applications. The reproducibility of this effect is confirmed by an additional sample shown in Supporting Information Note 4.

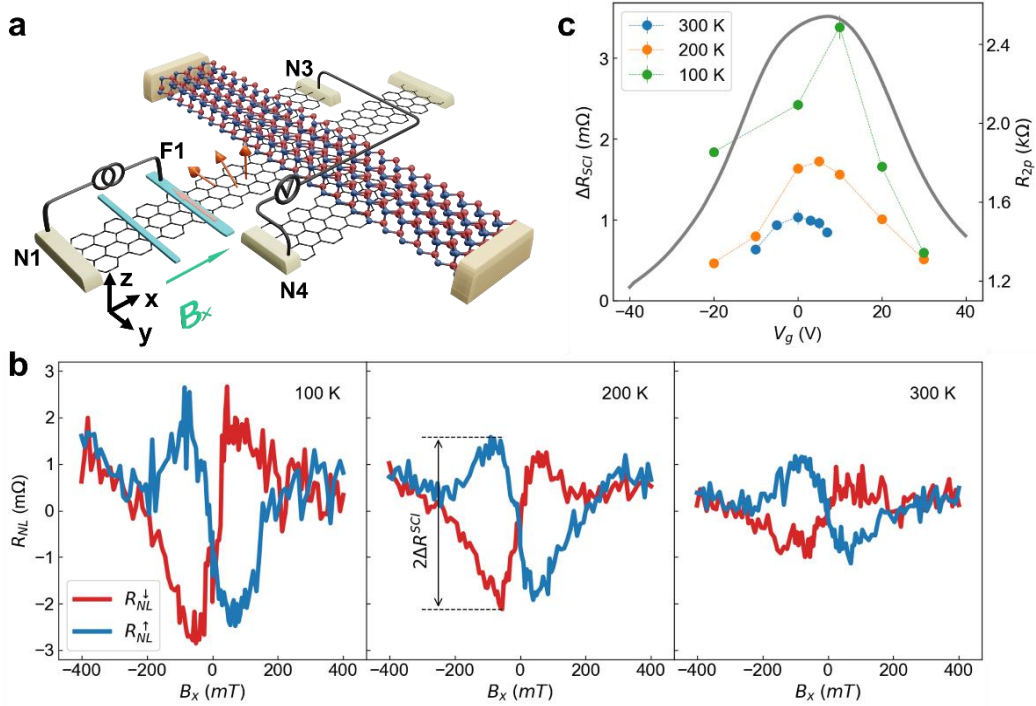

**Figure 4**. Spin-charge interconversion in proximitized graphene measured with spin precession. (a) Sketch of SCI measurement for antisymmetric Hanle precession at the SnS-proximitized graphene. $B_x$ is applied to induce precession of the *y*-polarized spins, which rotate towards the *z*-direction and are then converted into a charge current by the inverse SHE in the SnS-proximitized graphene. (b) Non-local resistance as a function of $B_x$ measured at 100, 200, and 300 K with $V_g$ = 0 V, with an initial positive ($R_{NL}^{\uparrow}$, blue line) and negative ($R_{NL}^{\downarrow}$, red line) magnetization direction of F1. The baseline is removed for a direct comparison of the SCI signal. The amplitude $\Delta R_{SCI}$ is labeled. (c) $\Delta R_{SCI}$ as a function of $V_g$ measured at 100, 200, and 300 K. The solid gray lines are the two-point resistance of the Hall arm, measured between N3 and N4 at 100 K.

Next, we study the gate tunability of the SCI. The $\Delta R_{SCI}$ amplitude of the antisymmetric Hanle curve (defined in Figure 4(b)) is plotted in Figure 4(c), with the raw data shown in Supporting Information Note 3. The error bars are calculated using the standard deviation of the antisymmetric component of the Hanle precession signal within the saturation region at the highest applied fields, where the signal should be constant and allows for a consistent quantification of the noise in the measurement. The two-point resistance between N3 and N4 measured at 100 K is also plotted as a gray solid line. Here, we see that the SCI for graphene/SnS heterostructure is gate tunable for all measured temperatures. $\Delta R_{SCI}$ shows the highest value around the CNP without changing sign, a phenomenon similarly reported in graphene/TMD[18,50] and graphene/metal oxide[30] heterostructures. When the temperature increases from 100 to 300 K, the peak values of $\Delta R_{SCI}$ shift towards more negative gate voltage. This trend mirrors the observations in Figure 2(d), where the CNP is seen to shift in the same direction due to the possible charge transfer between SnS and graphene.

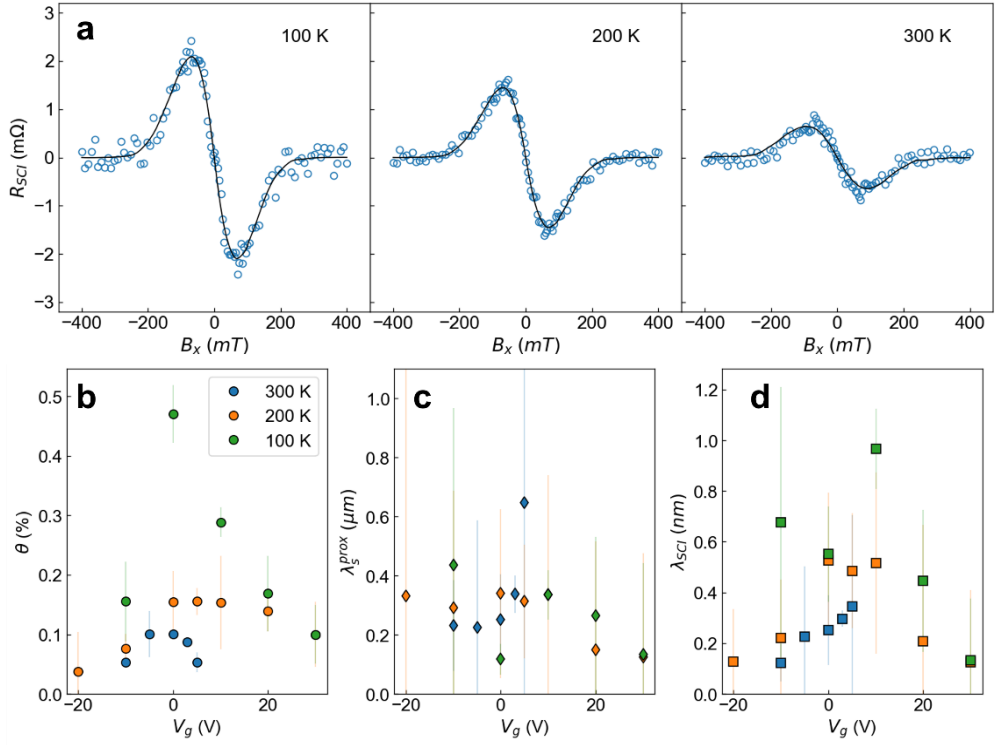

**Figure 5**. Gate tunability of the spin Hall angle and SCI efficiency. (a) Pure SCI signal obtained with $V_g$ = 0 V at 100, 200, and 300 K by subtracting the measured $R_{NL}$ curves between the positive and negative alignment of the F1 magnetization (shown in Figure 4(a)), followed by an antisymmetrization. The fitting at each temperature, using the solution of a Bloch function, is plotted as a solid black line. Extracted (b) spin Hall angle, (c) spin diffusion length of the proximitized graphene, and (d) SCI efficiency as a function of $V_g$ measured at 100, 200, and 300 K. The error bar indicates the fitting error of each parameter to the antisymmetric Hanle precession curve.

We then perform a quantitative analysis of the SCI efficiency. The pure SCI signal ($R_{SCI}$) with out-of-plane spin polarization is obtained by subtracting the $R_{NL}$ curves between the positive ($R_{NL}^{\uparrow}$) and negative ($R_{NL}^{\downarrow}$) alignment of the F1 magnetization, $R_{SCI} = (R_{NL}^{\uparrow} - R_{NL}^{\downarrow})/2$, followed by an antisymmetrization of the obtained curve with respect to the magnetic field. This process removes any initial-magnetization-independent components, such as local magnetoresistance, ordinary Hall effect, and Rashba-Edelstein effect[6,12,17,30]. The pure SCI signals obtained from the data in Figure 4(b) are shown in Figure 5(a), with the fitting curves plotted as solid black lines. The details of the fitting procedure are explained in Supporting Information Note 5.

In order to decrease the number of fitting parameters, we assume $D_s$ of the pristine graphene and the SnS-proximitized graphene to be equal. The relevant parameters extracted from the fitted data as a function of $V_g$ at different temperatures are shown in Figs. 5(b), (c) and (d). From the plot in Figure 5(b), we find that $\theta_{SH}$ exhibits a maximum around the CNP of the proximitized graphene, which is similar to the graphene/TMD heterostructure[18,50], as it has been calculated[25] that the spin Hall conductivity peaks around the CNP. When increasing the temperature, $\theta_{SH}$ decreases from 0.47±0.05% at 100 K to 0.10±0.01% at 300 K with $V_g$ = 0 V. The spin diffusion length of the proximitized region $\lambda_s^{prox}$ is shown in Figure 5(c). Due the SOC proximity, $\lambda_s^{prox}$ decreases from 2190±20 nm of the pristine graphene to 330±90 nm of the proximitized graphene at 100 K with $V_g$ = 10 V. The gate-dependent SCI efficiency $\lambda_{SCI}$, defined as the product of the spin Hall angle and the spin diffusion length of the proximitized region $(\theta_{SH} \cdot \lambda_s^{prox})$, is plotted in Figure 5(d). At 100 K and $V_g$ = 10 V, $\lambda_{SCI}$ is 0.97±0.16 nm. This number is larger than those for heavy metals (0.2 nm for Pt[51], 0.34 nm for W[52]) and metallic interfaces (0.3 nm for Ag/Bi[53], 0.17 nm for Cu/Au[54]). Whereas the SCI efficiency of SnS/graphene is smaller than that of graphene/TMD[14,16], it is similar that found in graphene/metal oxide heterostructures (1 nm for graphene/BiO$_x$[15] and 2 nm for graphene/CuO$_x$[30]). At 300 K, $\lambda_{SCI}$ decreased to 0.35±0.41 nm at $V_g$ = 5 V.

## 3. Conclusion

In conclusion, we report the first experimental demonstration of the SHE in trilayer graphene/group-IV monochalcogenide (SnS) heterostructures. The graphene acquires SOC from SnS by the proximity effect which leads to SHE up to room temperature. The SCI signal amplitudes are gate tunable, with a maximum value around the CNP of the SnS-proximitized graphene. Upon conducting a careful analysis considering the dependence of the spin injection efficiency and spin transport of graphene with gate voltage, we found both $\theta_{SH}$ and the SCI efficiency to be also tunable by the gate voltage. These parameters are found to be competitive with those observed in graphene/metal oxide heterostructures. Our study offers a novel approach for inducing SOC in graphene, enhancing its utility in spintronic applications. This advancement signifies a step forward in the manipulation and understanding of spin-based phenomena in 2D materials, potentially broadening the scope of graphene's applications in advanced electronic devices.

## 4. Experimental Section/Methods

*SnS crystal growth:* SnS was synthesized by direct reaction of elements in a quartz ampoule. Sulfur (99.9999%, 2-6 mm granuels, *Wuhan Xinrong* New Material Co., China) and tin (2-4 mm granules, 99.9999%, *Wuhan Xinrong* New Material Co., China) were placed in a quartz ampoule (25×160 mm) in stochiometric ratio corresponding to 25 g and melt-sealed under high vacuum (<7.5×10$^{-6}$ Torr, diffusion pump with LN$_2$ trap) using oxygen-hydrogen welding torch. The ampoule was placed horizontally in a muffle furnace and heated to 920°C with a heating rate of 0.5°C/min. After 6 h at this temperature, the ampoule was mechanically shaken in the furnace, cooled down to 850 °C using a cooling rate of 6°C/h, and subsequently to room temperature at 1°C/min. The formed chunk consisting of crystals with length exceeding 1 cm were removed from the ampoule in an Ar-filled glovebox.

*Device fabrication:* SnS flakes are produced by mechanical exfoliation of the SnS crystal onto polydimethylsiloxane (Gelpak PF GEL film WF 4, 17 mil.) inside an Ar-filled glovebox with O$_2$ concertation below 1.0 ppm and H$_2$O below 0.5 ppm. Graphene flakes are exfoliated onto 300-nm-thick SiO$_2$ on a doped Si substrate in atmosphere with Scotch tape, and then transferred into the glovebox immediately. The desired flakes are identified using an optical microscope. The selected SnS flake is then stamped on top of the graphene flake using a viscoelastic stamping tool in which a three-axis micrometer stage is used to position the flake accurately. After stamping, the graphene flake is nanopatterned into Hall bars, using electron-beam lithography, and covered with a 20-nm-thick Al deposited by thermal evaporation (base pressure ~1×10$^{-6}$ Torr) as a hard mask, followed by reactive-ion etching with Ar/O$_2$ plasma. The Al hard mask is then chemically removed with tetra-methyl ammonium hydroxide. Subsequently, the sample is annealed at 280°C for 60 minutes in ultrahigh vacuum (~1.5×10$^{-8}$

Torr) to remove fabrication residues. The Hall bars are then contacted with Pd(5 nm)/Au(45 nm) by electron-beam evaporation and lift-off in acetone. After fabricating the non-magnetic contacts, the magnetic $TiO_x$/Co electrodes are fabricated by e-beam lithography, deposition of 2.7 Å of Ti (followed by oxidation in air for 10 minutes), 35 nm of Co, and 15 nm of Au as a capping layer by e-beam lithography, evaporation, and lift-off. The width of the magnetic electrodes are 150 nm and 300 nm in order to have different coercivity. The presence of $TiO_x$ between the Co electrode and the graphene channel leads to interface resistances between 1.5 and 4 kΩ, measured with a conventional three-point measurement. The optical images during the fabrication process can be found in Supporting Information Note 1.

*Device characterization:* The exact dimensions of the devices are extracted from atomic force microscopy images (10 x 10 μm and 50 x 50 μm) obtained after the electrical measurements. The AFM images are acquired in tapping mode with Agilent 5500, Keysight scanning probe microscope using NCHR-50 tips with 10N/m cantilever's elastic constant. Micro-Raman spectroscopy measurements are carried out at room temperature after the electrical characterization, by placing the samples in a Linkam® vacuum chamber (~$3\times10^{-5}$ Torr) coupled to a Renishaw® inVia Qontor Raman instrument equipped with a 50× objective (Nikon®, N.A. 0.60, WD 11 mm). We use a 532 nm laser as excitation source with a 2400 l/mm grating. The laser power was kept < 0.5 mW to avoid damaging the samples during the measurement.

*Electrical measurements:* The electrical measurements are performed in a Physical Property Measurement System (PPMS) by Quantum Design, using a DC reversal technique with a Keithley 2182 nanovoltmeter and a 6221 current source. The n-doped Si substrate acts as a back-gate electrode to which we apply the gate voltage with a Keithley 2636B. The semiconducting properties of the SnS flake are measured with a Keithley 2636B, which utilize two individual channels for controlling the source-drain and gate voltages, while monitoring the source-drain and leakage currents, respectively. The leakage current is below 1 nA during all gate-dependent measurements reported.

## Supporting Information

Supporting Information is available from the Wiley Online Library.

## Acknowledgements

Haozhe Yang, Zhendong Chi and Garen Avedissian contributed equally to this work. We acknowledge funding by the "Valleytronics" Intel Science Technology Center, by the Spanish MICIU/AEI/10.13039/501100011033 and by ERDF A way of making Europe (Project No. PID2021-122511OB-I00 and "Maria de Maeztu" Units of Excellence Programme No. CEX2020-001038-M), and by Diputación de Gipuzkoa (Project No. 2021-CIEN-000037-01). This project has received funding from Marie Sklodowska-Curie Actions, H2020-MSCA-ITN-2020; Project Acronym SPEAR; Grant Agreement No. 955671. Z.C. acknowledges postdoctoral fellowship support from "Juan de la Cierva" Programme by the Spanish MICIU/AEI (grant No. FJC2021-047257-I). H .Y thanks support from NSFC 92164206 and 52261145694. B. M.-G. thanks support from "Ramón y Cajal" Programme by the Spanish MICIU/AEI (grant No. RYC2021-034836-I). CzechNanoLab project LM2023051 funded by MEYS CR is gratefully acknowledged for the financial support of the measurements at LNSM Research Infrastructure. Z.S. was supported by ERC-CZ program (project LL2101) from Ministry of Education Youth and Sports (MEYS) and by the project Advanced Functional Nanorobots (reg. No. CZ.02.1.01/0.0/0.0/15_003/0000444 financed by the EFRR).

## Competing interests

The authors declare no competing interests.

## Data and code availability

All the data and codes used in this study are available from corresponding authors upon reasonable request.

Supporting Information

**Note 1. Optical image of Sample 1 during the fabrication process.**

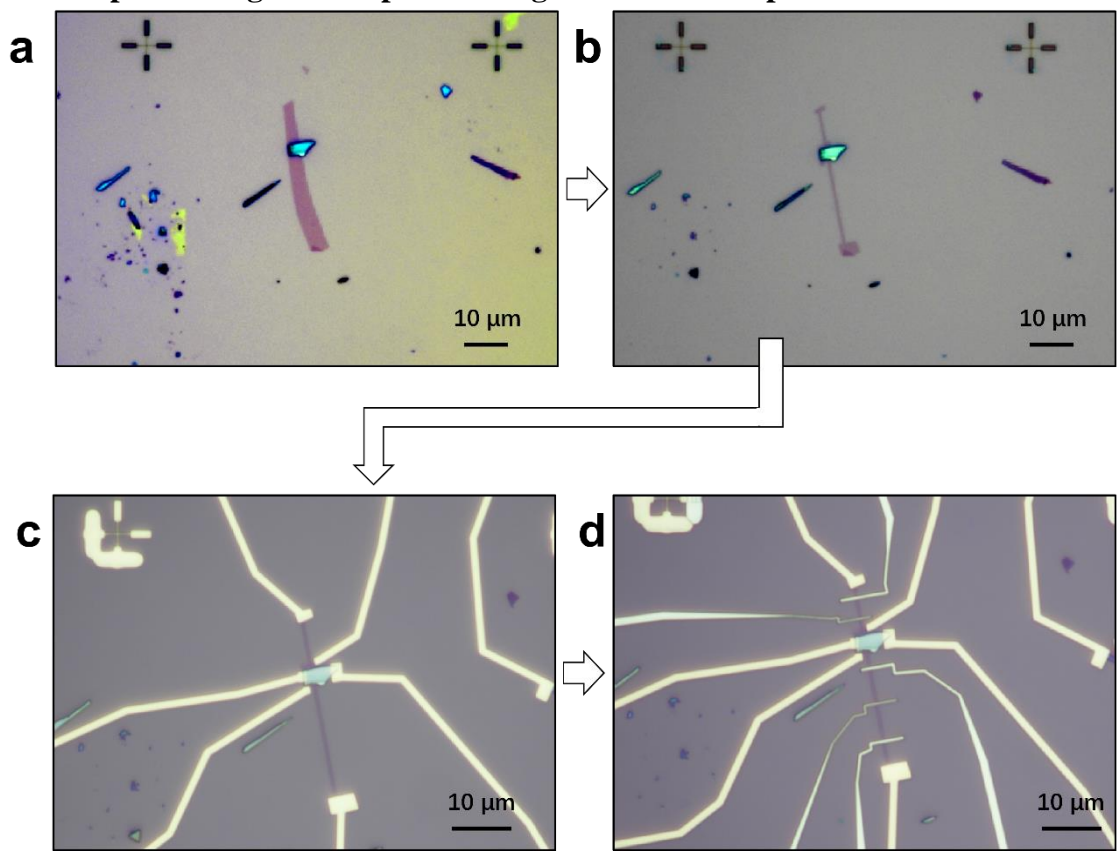

**Figure S1** (a) Stamped SnS/trilayer graphene on the substrate; (b) Graphene Hall bar after reactive ion etching; (c) Non-magnetic Pd/Au contacts deposition; (d) Ferromagnetic TiO$_x$/Co contacts deposition.

**Note 2. Raw data of symmetric Hanle precession.**

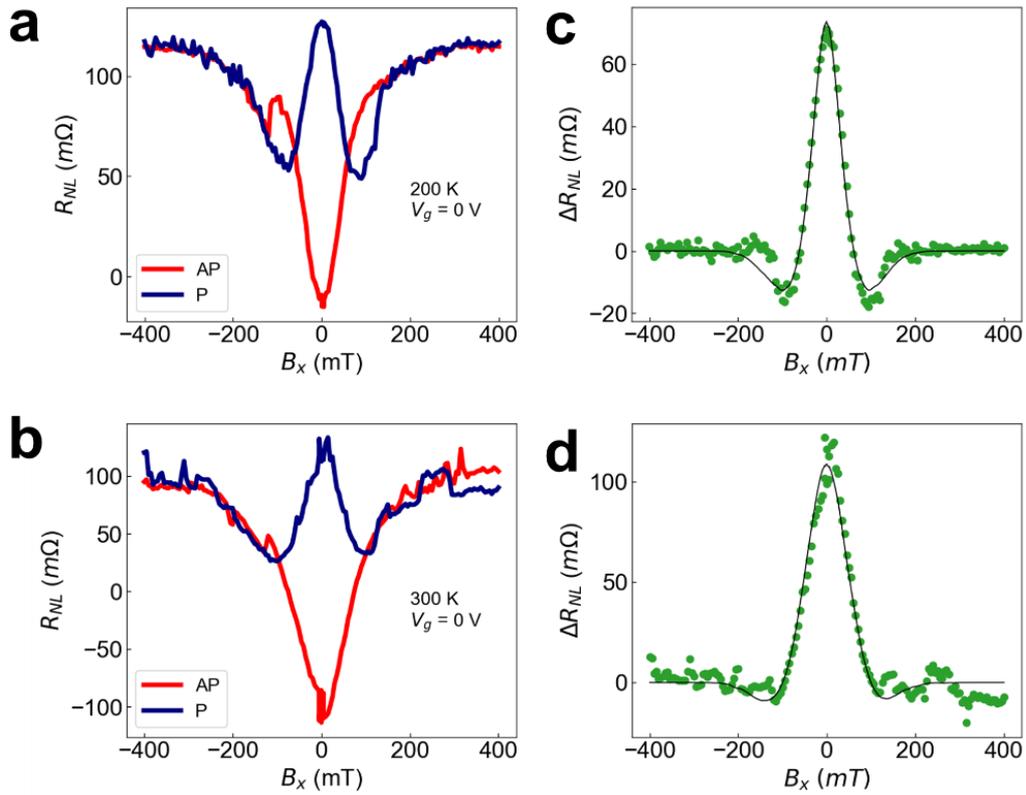

**Figure S2** Hanle precession measurement measured at (a) 200 K and (b) 300 K and $V_g = 0$ V, with the measurement configuration shown in Figure 3a. (c, d) Net symmetric Hanle precession signal extracted from the two curves in (a, b), respectively. The gray solid line is a fit of the data to the solution of the Bloch equation.

**Note 3. Raw data of the gate dependence of the SCI conversion.**

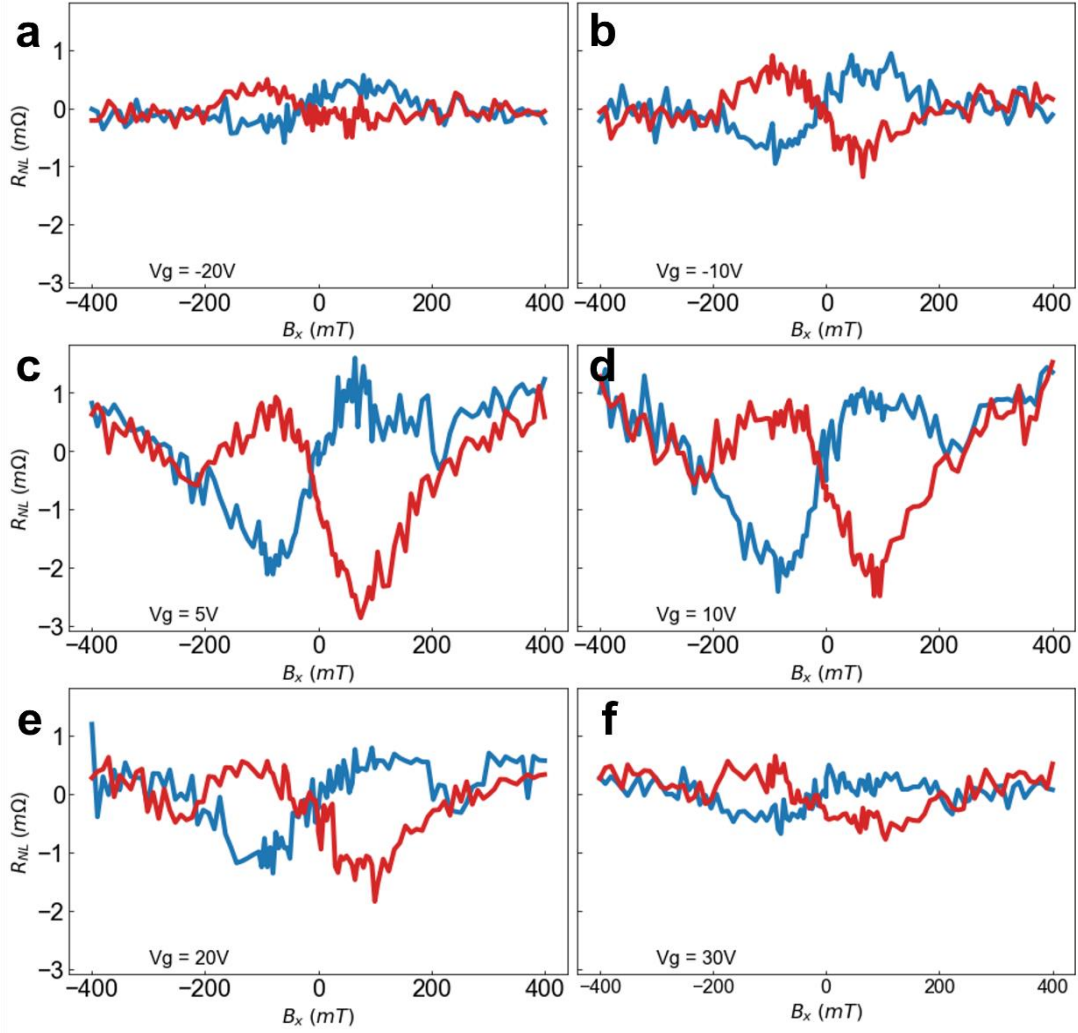

**Figure S3** Nonlocal resistance as a function of $B_x$ measured at 200 K and gate voltage $V_g$ of (a) −20 V, (b) −10 V, (c) 5 V, (d) 10 V, (e) 20 V, and (f) 30 V, respectively. The measurement configuration is shown in Figure 4a. The magnetization direction of F1 is initialized positive ($R_{NL}^\uparrow$, blue line) and negative ($R_{NL}^\downarrow$, red line). The baseline is removed for a direct comparison of the SCI signal.

**Note 4. Reproducibility.**

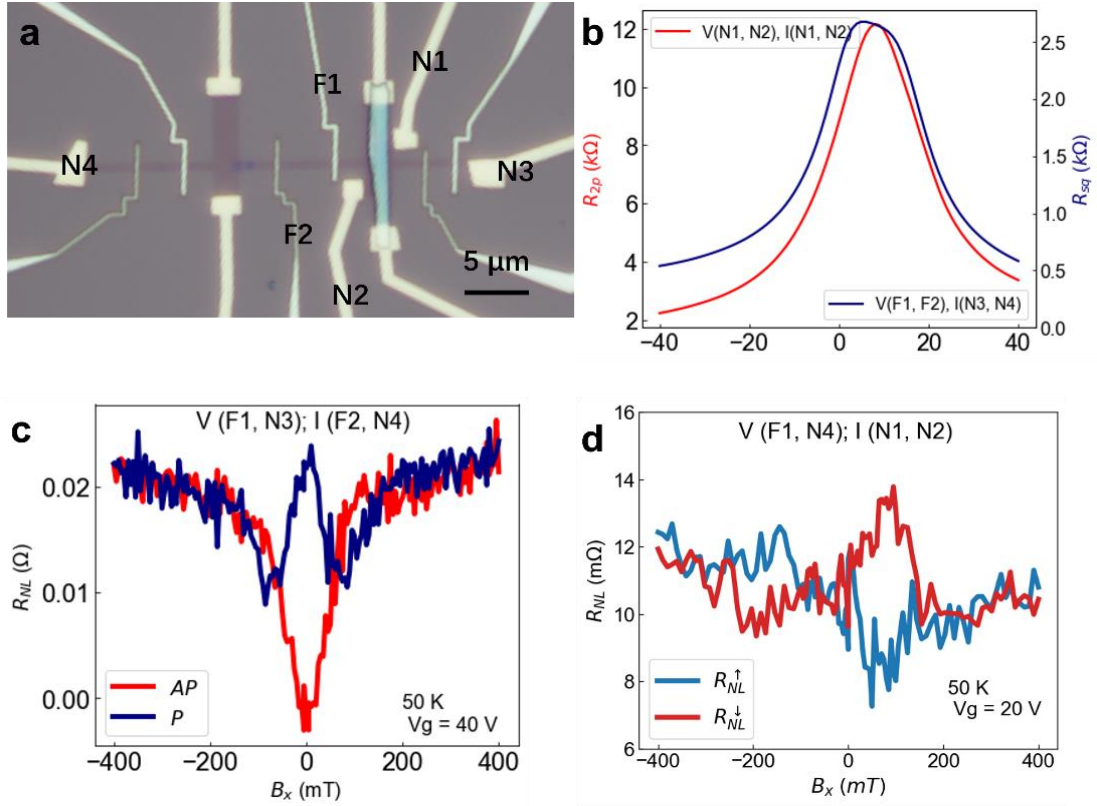

**Figure S4** (a) Optical image of Sample 2. (b) Square resistance of pristine graphene (blue solid line) and 2-point resistance of the SnS-covered graphene at the cross-junction of the Hall bar (red solid line) measured in Sample 2. (c) Non-local resistance as a function of $B_x$ measured at 50 K and $V_g$ = 40 V, with F1 and F2 electrodes set in a parallel ($R_{NL}^P$, blue line) and antiparallel ($R_{NL}^P$, red line) configuration. The measurement configuration is labeled in the panel. (d) Non-local resistance as a function of $B_x$ measured at 50 K and $V_g$ = 20 V, with initial positive ($R_{NL}^\uparrow$, blue line) and negative ($R_{NL}^\downarrow$, red line) magnetization direction of F1. The measurement configuration is labeled in the panel.

**Note 5. Spin precession fitting process.**

To analyze the symmetric Hanle precession experiments obtained while having a parallel and or antiparallel orientation of the Co magnetization, we model the spin propagation in our devices using the Bloch equations:

$$D_s \nabla^2 \vec{\mu} - \frac{\vec{\mu}}{\tau_s} + \vec{\omega} \times \vec{\mu} = 0, \quad (1)$$

where $\vec{\mu}$ is the spin accumulation, $D_s$ the spin diffusion constant, and $\tau_s$ the spin lifetime. $\vec{\omega} = g\mu_B \vec{B}$ is the Larmor frequency, $g=2$ is the Landé factor, $\mu_B$ is the Bohr magneton, and $\vec{B}$ is the applied magnetic field.

The spin precession is induced in the $y - z$ plane when applying a magnetic field along $x$ direction. In this case, Eq. (1) turns into

$$D_s \frac{d^2}{dx^2} \begin{pmatrix} \mu_x \\ \mu_y \\ \mu_z \end{pmatrix} - \begin{pmatrix} \tau_x^{-1} \\ & \tau_y^{-1} \\ & & \tau_z^{-1} \end{pmatrix} \begin{pmatrix} \mu_x \\ \mu_y \\ \mu_z \end{pmatrix} + \vec{\omega} \begin{pmatrix} 0 \\ \mu_y \\ -\mu_z \end{pmatrix} = 0. \quad (2)$$

The solution of $\mu_y$ and $\mu_z$ in Eq. (2) is

$$\mu_y = A e^{\frac{x}{\lambda_s}\sqrt{1+i\omega\tau_s}} + B e^{\frac{x}{\lambda_s}\sqrt{1-i\omega\tau_s}} + C e^{-\frac{x}{\lambda_s}\sqrt{1+i\omega\tau_s}} + D e^{-\frac{x}{\lambda_s}\sqrt{1-i\omega\tau_s}} \quad (3)$$

$$\mu_z = -iA e^{\frac{x}{\lambda_s}\sqrt{1+i\omega\tau_s}} + iB e^{\frac{x}{\lambda_s}\sqrt{1-i\omega\tau_s}} - iC e^{-\frac{x}{\lambda_s}\sqrt{1+i\omega\tau_s}} + iD e^{-\frac{x}{\lambda_s}\sqrt{1-i\omega\tau_s}} \quad (4)$$

where $\lambda_s = \sqrt{\tau_s D_s}$ is the spin diffusion length. $A$, $B$, $C$ and $D$ are the coefficients determined by the boundary conditions. The spin current is defined as:

$$I_{Sy(z)} = -\frac{W_{gr}}{eR_{sq}} \frac{d\mu_{y(z)}}{dx}, \quad (5)$$

where $W_{gr}$ is the width of the graphene channel and $R_{sq}$ is the square resistance of graphene. The spin accumulation at the detection position $x = x_{det}$ is then converted into a voltage with:

$$V_{det}^S = P \frac{\mu_y(x_{det})}{e}, \quad (6)$$

where $P$ is the spin polarization of the Co detector. We consider that the Co injector has the same spin polarization. $V_{det}^S$ is usually normalized by the charge current $I_c$, giving the non-local resistance:

$$R_{NL} = \frac{V_{det}^S}{I_c}. \quad (7)$$

The non-local resistance is also affected by the contact pulling effect[1]. Because of the finite/small in-plane shape anisotropy of the ferromagnetic electrodes, the magnetization is then pulled by an angle β from their easy axis towards the field direction, injecting spins along $x$ and resulting in an additional term to the non-local resistance. The $R_{NL}$ obtained in this case is:

$$R_{NL} = \pm R^s cos(β_1)cos(β_2) + R^{s0} sin(β_1)sin(β_2). \qquad (8)$$

Here ± represents the non-local signal with parallel (P) and antiparallel (AP) configuration of the Co electrode magnetizations. $β_1$ and $β_2$ correspond to the angle β of the two magnetic electrodes used as injector and detector. For simplicity, we consider that the two Co electrodes have the same pulling effect with the angle, $β = β_1 = β_2$. $R^{s0}$ is a constant that corresponds to the spin signal at zero magnetic field. The net spin signal $\Delta R_{NL}$ is then expressed as:

$$\Delta R_{NL} = \frac{R_{NL}^P - R_{NL}^{AP}}{2} = R^s \cos^2(β). \qquad (9)$$

Because of the opposite spin precession sign of $R_{NL}^P$ and $R_{NL}^{AP}$, the sum $R_{sum} = R_{NL}^P + R_{NL}^{AP}$ is then proportional to $\sin^2(β)$. For our analysis, we used the contact pulling from the measurement points on the curve of $R_{sum}$ vs. $B_x$, with a representative result of Sample 1 shown in Figure S5 measured at 200 K and $V_g = 0$ V.

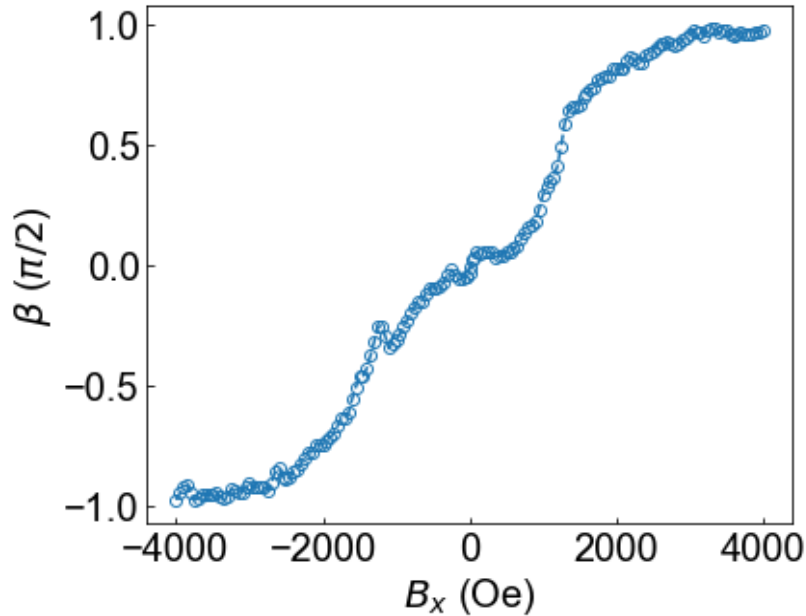

**Figure S5** Contact pulling factor β as a function of the in-plane hard axis magnetic field $B_x$.

We determine the spin signal using the following boundary conditions:
1. Spin accumulation μ is continuous.
2. Spin current is continuous except:

a. At F1 with spin injection by the charge current, by $\Delta I_s = I_c \cdot P/2$;
b. At F1 and F2 due to the spin backflow effect because of the low contact resistance, by $\Delta I_s = -\mu_s/(2eR_c)$, where $R_c$ is the contact resistance.
3. Spin current is zero at the sample ends.

We write these boundary conditions in a matrix $X$ which fulfills: $MX = Y$, where $M$ contains the coefficients $M = A, B, C \ldots$ and $Y = (0, \ldots, \frac{I_c P}{2}, 0, \ldots)$ and use the Moore-Penrose inverse to invert $X^{-1}$ and obtain $M = YX^{-1}$. Using the numeric solution described above, we input the measured net spin signal $\Delta R_{NL}$, square resistance $R_{sq}$, pulling factor $\cos^2(\beta)$, channel size $W_{gr}$ determined by SEM, and the contact resistance $R_c$.

We determine the spin Hall angle $\theta_{SH}$ and the spin lifetime $\tau_s^{prox}$ of the graphene in the heterostructure with the antisymmetric spin precession signal $R_{SCI}$ originated from the (inverse) spin Hall effect of functionalized graphene. Here, we derive the inverse SHE but, due to reciprocity, the obtained equation is valid for both configurations. The non-local resistance $R_{SCI}$ is given by

$$R_{SCI} = \frac{\theta_{SH} R_{sq} \overline{I_{Sz}}}{I_c} \cdot \cos(\beta) \qquad (10)$$

where $\overline{I_{Sz}}$ is the average spin current in the cross-junction area of the Hall bar, which is defined as $\overline{I_{Sz}} = \frac{1}{W_{cr}} \int_L^{L+W_{cr}} I_{Sz}(x)dx$, where $L$ is the distance from F1 electrode to the edge of the Hall cross and $W_{cr}$ is the width of the cross. $\cos(\beta)$ is the term indicating the contact pulling of the ferromagnetic electrode. Note that, since there is only one magnetic electrode, this term is not squared as in Eq. (9). To determine $I_{Sz}(x)$, we use the equations and boundary conditions described above. Finally, since the CNP is the same for the pristine and proximitized regions, we assume the spin diffusion coefficients for the proximitized region and the pristine region to be equal.

**Note 6. Atomic force microscopy of the device.**

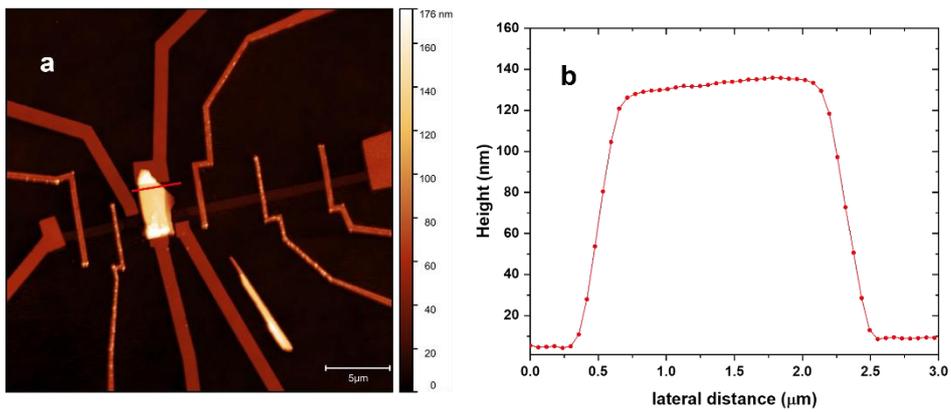

**Figure S6** (a) AFM image of Sample 1 after the electrical measurement. (b) Line-scan profile of along the red solid line in (a), showing the thickness of the SnS flake is about 120 nm.